\documentclass[default,iicol]{sn-jnl}% Default with double column layout

\usepackage{graphicx}%
\usepackage{multirow}%
\usepackage{amsmath,amssymb,amsfonts}%
\usepackage{amsthm}%
\usepackage{mathrsfs}%
\usepackage[title]{appendix}%
\usepackage{xcolor}%
\usepackage{textcomp}%
\usepackage{manyfoot}%
\usepackage{booktabs}%
\usepackage{algorithm}%
\usepackage{algorithmicx}%
\usepackage{algpseudocode}%
\usepackage{listings}%
\newcommand{\onlinecite}[1]{\hspace{-1 ex} \nocite{#1}\citenum{#1}}
\begin{document}
%%%%%%%%%%%%%%%%%%%%%%%%%%%%%%%%%%%%%%%%%%%%%%%%%%%%%%%

\title[Article Title]{\vspace{-1cm}Optical spectra of silver clusters and nanoparticles of all sizes from the TDDFT+U method}

\author*[1,2]{\fnm{Mohit} \sur{Chaudhary}}\email{mohit.chaudhary@univ-amu.fr}

\author*[1,2]{\fnm{Hans-Christian} \sur{Weissker}}\email{hans-christian.weissker@univ-amu.fr}

\affil[1]{\orgname{Aix-Marseille University, CNRS, CINaM UMR 7325}, \city{13288 Marseille}, \country{France}}

\affil[2]{\orgname{European Theoretical Spectroscopy Facility (ETSF)}}

%%==================================%%
%%               abstract           %%
%%==================================%%

\abstract{The localized surface-plasmon resonances (LSPRs) of coinage-metal clusters and nanoparticles provide the basis for a great number of applications, the conception and necessary optimization of which require precise theoretical description and understanding. However, for the size range from clusters of a few atoms through nanoparticles of a few nanometers, where quantum effects and atomistic structure play a significant role, none of the methods employed to date has been able to provide high-quality spectra for all sizes. The main problem is the description of the filled shells of d electrons which influence the optical response decisively. In the present work we show that the DFT+U method, employed with real-time time-dependent density-functional theory calculations (RT-TDDFT), provides spectra in good agreement with experiment for silver clusters ranging from 4 to 923 atoms, the latter representing a nanoparticle with a diameter of 3\,nm. Both the electron-hole-type discrete spectra of the smallest clusters and the broad plasmon resonances of the larger sizes are obtained. All calculations have been carried out using the same value of the effective $U$ parameter that has been found to provide good results in bulk silver. The agreement with experiment for all sizes shows that the $U$ parameter is surprisingly transferable. Our results open the pathway for TDDFT calculations of many practically relevant systems including clusters coupled to bio-molecules or to other nano-objects.}


\maketitle

%%%%%%%%%%%%%%%%%%%%%%%%%%%%%%%%%%%%%%%%%%%%%%%%%%%%%%%

\section{Introduction}\label{sec1}
Noble-metal clusters and nanoparticles are employed in an overwhelming number of applications and research domains~\cite{schwartzberg2008novel,chakraborty-pradeep-review-17,gilroy-16,zhang-review-16,mathew-pradeep-14}. In particular, there is enormous interest in their optical properties, mostly connected to the localized surface-plasmon resonances (LSPRs) and their tuning and application, which creates a natural link to the field of nanoplasmonics and quantum plasmonics~\cite{tame-13,marinica-12,varas-16}. The coupling of metal clusters with, for example, organic molecules, the exploitation of field enhancements between clusters~\cite{tripathy2011acousto} or around edges and tips~\cite{hao2007plasmon} as in surface-enhanced Raman spectroscopy~\cite{zong-SERS-18}, and the applications of the clusters as sensors~\cite{mejia2018plasmonic} call for a predictive and very precise theoretical description. However, a longstanding problem has been exactly the precise description of the coinage-metal clusters' optical spectra and surface-plasmon resonance energies, necessary in order to model and analyze the interaction for instance with biomolecules.

Any description of such interactions clearly needs to take into account all the quantum effects at play as well as the effects of the atomic structure and of chemical bonds present in the systems.

%%%%%%%%%%%%%%%%%%%%%%%%%%%%%%%%%%%%%%%%%%%%%%%%%%%%%%%

The well-known principal problem for the coinage metals is the proper description of the filled d shell of electrons. Interband transitions from the d electrons into states above the Fermi energy appear in the spectra of bulk silver at about 4\,eV, and of gold at about 2\,eV~\cite{johnson-72}. In addition, in the presence of a LSPR, the d electrons are polarized inside the material by the field created by the collective oscillation of the delocalized electrons, opposing the latter~\cite{liebsch-93,PRL-serra-rubio-97,cottancin2006optical,weissker-real-time-15,sinha-roy-23}. This leads to a screening which shifts the LSPR energy to lower energies~\cite{cottancin2006optical}. The opposite polarizations are easily seen in the induced densities at the plasmon energy~\cite{weissker-real-time-15,sinha-roy-23}. The d electrons are strongly localized around the atom cores, unlike the delocalized s electrons that produce the spill-out over the classical particle radius that produces a red-shift of LSPR energies~\cite{liebsch-93b,kim-97,liebsch-98}. This leads to the concept of \textit{a layer of reduced screening of the d electrons} at the surface of the clusters. The interplay of these two effects determines the size-dependence of LSPR energies of clusters in vacuum~\cite{cottancin2006optical,haberland-13,campos2019plasmonic}.
 \\
  
%%%%%%%%%%%%%%%%%%%%%%%%%%%%%%%%%%%%%%%%%%%%%%%%%%%%%%%

For ``large'' nanostructures, purely classical approaches do well~\cite{jensen1999electrodynamics,kelly2003optical,maier-book-07} (Mie theory in the case of spherical nanoparticles). For smaller particles, hydrodynamic~\cite{christensen2014nonlocal} and other non-local classical approaches~\cite{sinha2017classical} have tried to include at least part of the relevant surface and quantum effects with some success. To include quantum effects for intermediate sizes, jellium-based calculations of Time-Dependent Density-Functional Theory (TDDFT) obtain excellent results for, e.g., the size dependence of LSPR energies, but at the price of ignoring atomistic structure and interfacial details~\cite{cottancin2006optical,campos2019plasmonic}. The smallest clusters, however, are clearly quantum systems with discrete electronic states which necessitate a full atomistic quantum description. Early on, very small clusters have been described by quantum-chemistry methods like the equation-of-motion coupled-cluster approach~\cite{vlasta-01}. 

Today, TDDFT (using pseudopotentials or similar descriptions of the electron-ion interaction) has become the workhorse of calculations on clusters, but it hasn't achieved, until now, the predictive quality over the full size range of clusters and nanoparticles that is needed. In particular, the approximations involved in the description of exchange and correlation effects (functionals and kernels), might be well adapted to strongly localized systems where short-range effects play an important role, or else to more extended systems where long-range interactions are important~\cite{sottile-IJQC,weissker-ixs-long}. Over the full size range interesting for the clusters and nanoparticles, we have both regimes, which in addition might be combined in the same system, as in the case of active tips and edges of a cluster, of tiny clusters attached to larger NPs, or in the interaction of nanoparticles with molecules.

For small silver clusters, the importance of long-range exchange effects has been shown~\cite{rabilloud2013assessment}, and range-separated hybrid functionals provide spectra in excellent agreement with experiment~\cite{anak2014time,schira2019localized}. However, these calculations are numerically cumbersome and today limited to sizes of up to $\approx$ 150... 200 atoms~\cite{schira2019localized,seveur-23}. In addition, the published results for one of the largest attainable clusters (Ag$_{147}$) seem to overestimate the LSPR energy with respect to available experiment (Ref.~\cite{schira2019localized}, cf.\ Fig.~\ref{fig_energies}).

For larger clusters, local and semilocal functionals like the simple local-density approximation (LDA)~\cite{iida-14} and different generalized-gradient approximations (GGA)~\cite{perdew1996generalized,WC-functional,LB94} have been the most widely used approximations until now~\cite{malola-13,sinha-roy-23,koval-16,aikens-08,burgess-14}, along with meta-GGA functionals for larger gold clusters~\cite{sakthivel-20-VS98} and the GLLB-SC functional (SC standing for ``solids and correlation'')~\cite{kuisma-10,GLLB-functional}. The latter seems to be well adapted to large clusters with a clear plasmonic resonance but does less well for sizes below about 100 atoms~\cite{kuisma2015localized}. In addition, a number of approximate schemes based on TDDFT have been developed in order to reduce the numerical effort and more easily attain larger systems, like the DFTB~\cite{magrebi-23} and approximate TDDFT algorithms~\cite{baseggio-15}.

The use of the ``simple'' functionals results, in general, in the filled d shells positioned too close to the Fermi energy and, due to the resulting overestimation of the screening of the LSPR by the d electrons, underestimates the LSPR energies~\cite{rabilloud-14,weissker2011optical,weissker2014optical,sinha-roy-23}.

%%%%%%%%%%%%%%%%%%%%%%%%%%%%%%%%%%%%%%%%%%%%%%%%%%%%%%%
What is needed is a method which selectively improves the description of the d electrons while avoiding the costly introduction of Hartree-Fock exchange as in the hybrid functionals. This can be achieved by the DFT+U method as introduced by Anisimov, Liechtenstein, and coworkers~\cite{anisimov-97,anisimov-93,liechtenstein-95,anisimov-91} which corrects DFT calculations for problems mostly related to the ``over-delocalization'' of the d electrons, resulting from the self-interaction problem that follows from the incomplete cancellation of the Coulombic terms when approximate density functionals are used. 

It has been demonstrated that for bulk metals, the DFT+U approach provides good dielectric functions that have then been used in the classical calculations of the optical response of large nanoparticles~\cite{avakyan2020theoretical}. However, this approach raises the usual questions of transferability and of the validity of the very concept of the dielectric function for small clusters. Coviello \textit{et al.} have recently extended this approach to magnetic elements~\cite{coviello2024}. Explicit DFT+U calculations do not seem to have been published for noble-metal clusters.
\\

In the present work, we use the DFT+U approach~\cite{anisimov-97,anisimov-93,liechtenstein-95,anisimov-91} and its extension time-dependent DFT+U (TDDFT+U) to obtain spectra of silver clusters in good agreement with available experiments over the full size range spanning from few-atom clusters like Ag$_4$ through nanoparticles of about 1000 atoms (Ag$_{923}$ with more than 10,000 active electrons, about 3\,nm in diameter, presently certainly inaccessible for calculations with hybrid functionals). In particular, this includes the discrete electron-hole-type spectral features of the smallest clusters as well as the plasmonic response of larger ones, including the LSPR's complex size dependence and the oscillation-like behavior of small clusters due to shell-closing effects--- even though significant differences between different measurements exist, notably between low-temperature rare-gas-embedded clusters and gas phase measurements, which complicates the comparison and limits its precision. The calculated results are obtained using the same value of the effective $U$ for all sizes, showing surprising transferability of this parameter. The numerical effort of the DFT+U method is only slightly higher than that of the comparable ``pure'' DFT calculations.  Our results demonstrate that DFT+U is an \textit{efficient}
and \textit{transferable} method to model the electronic response of Ag clusters which will enable precise, predictive TDDFT calculations of many clusters, cluster+molecule hybrid systems, and cluster-assembled materials.

%%%%%%%%%%%%%%%%%%%%%%%%%%%%%%%%%%%%%%%%%%%%%%%%%%%%%%%
\section{Results}\label{sec2}
%%%%%%%%%%%%%%%%%%%%%%%%%%%%%%%%%%%%%%%%%%%%%%%%%%%%%%%

The DFT+U method corrects DFT calculations for problems mostly related to the ``over-delocalization'' of the d electrons, arising from the self-interaction errors due to the incomplete cancellation of the Coulombic terms when approximate density functionals are used. The main effect when applying the correction to the filled d states in silver is their downward shift with respect to the Fermi level and their increased localization.

In the present work, we apply the DFT+U method in its rotationally invariant formulation~\cite{dudarev1998electron} using the octopus code~\cite{tancogne2017self}, where the pseudo-wavefunctions serve as localized basis. DFT+U is used for the ground-state calculations and in the subsequent time-evolution (real-time, RT)~\cite{yabana1996time} TDDFT+U calculations, carried out to obtain the optical spectra~\cite{tancogne2017self}. We use a constant value of the effective $U$ of 4.0\,eV. This choice was motivated by the findings of Avakyan \textit{et al.}, who obtained a good agreement with the experimental dielectric function of \textit{bulk} silver using this value~\cite{avakyan2020theoretical}.
\\

%%%%%%%%%%%%%%%%%%%%%%%%%%%%%%%%%%%%%%%%%%%%%%%%%%%%%%%
\begin{figure*}[t]
  \centering
  \includegraphics[width=1.0\textwidth]{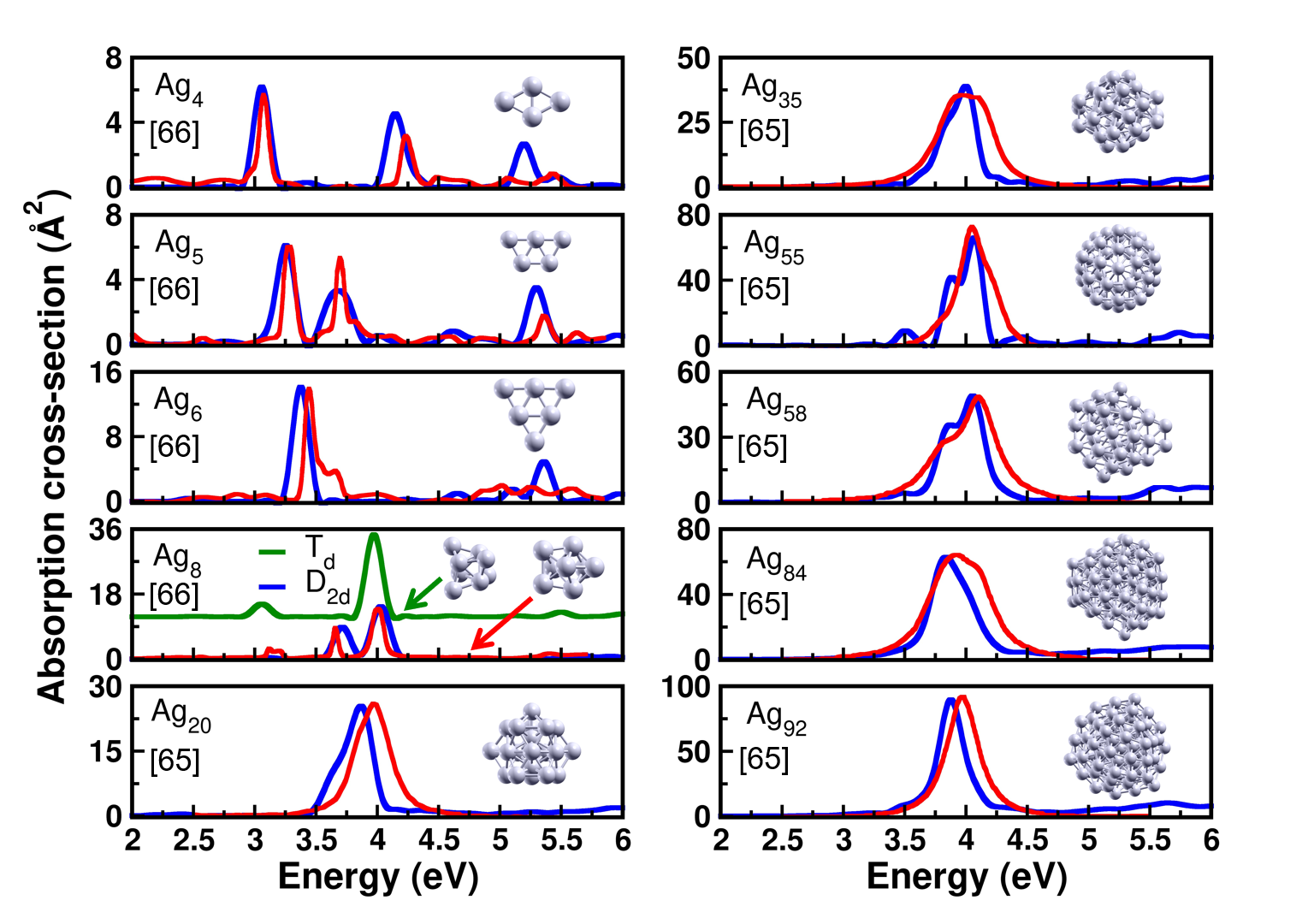}
  \caption{Absorption spectra of theoretically found lowest-energy structures of Ag$_n$ clusters, where n = 4$-$92. Blue curves represent the calculated absorption spectra with the TDDFT+U scheme with an effective $U$ value of 4\,eV. The red curves are the experimental optical absorption spectra scanned from the references~\protect\cite{yu2018optical,lecoultre2011ultraviolet} as indicated. Consequently, as already carried out in the references, the experimental curves corresponding to the clusters Ag$_{20}$ through Ag$_{92}$ are blue-shifted by 0.17\,eV to account for the neon matrix dielectric effect. For a detailed discussion of these shifts and their impact on our comparisons, please refer to section 1 of the supplementary material and Fig. S1. In the case of Ag$_8$, it appears likely that two different isomers coexist in the sample which need to be included in the calculation in order to reproduce the experimental peaks, as already pointed out in Refs.~\protect\onlinecite{lecoultre2011ultraviolet} and \protect\onlinecite{harb-08}.}
    \label{fig_spectra}
\end{figure*}
%%%%%%%%%%%%%%%%%%%%%%%%%%%%%%%%%%%%%%%%%%%%%%%%%%%%%%%

%%%%%%%%%%%%%%%%%%%%%%%%%%%%%%%%%%%%%%%%%%%%%%%%%%%%%%%
\begin{figure*}
\centering
\includegraphics[width=1.0\textwidth]{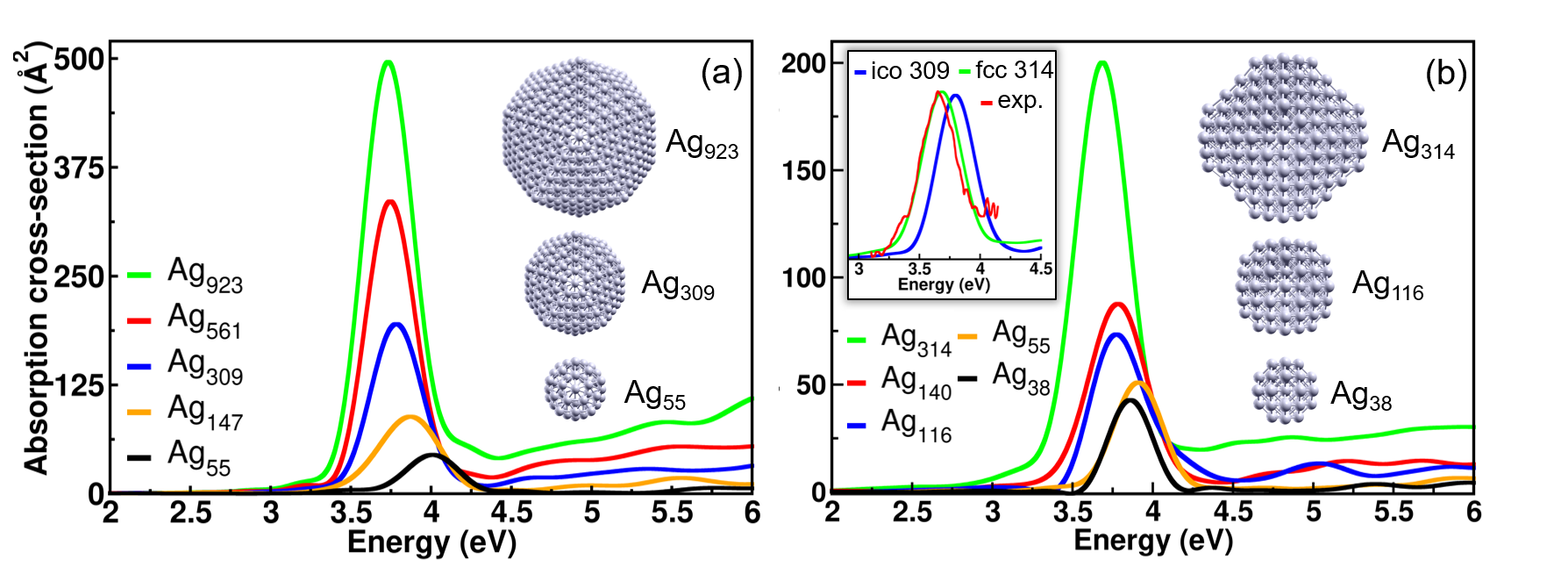}

\caption{Calculated absorption spectra of (a) Ag$_n$ icosahedral clusters, with n ranging from 55 to 923 atoms, and (b) fcc-based nearly spherical clusters ranging from 38 to 314 atoms (note the different scales). With decreasing size, the LSPR energy of the icosahedral clusters shows a smooth, monotonous increase from 3.73\,eV for the biggest cluster to 4.01\,eV for the smallest. As their shapes are equivalent, the changes are essentially a size effect. By contrast, for the fcc-based clusters, the shapes of the different sizes cannot all be equivalent, which results in different degrees of deviation from the ideal of a spherical cluster and, in turn, from a smooth, monotonous change of the plasmon energies. Inset in panel (b): spectra of the icosahedral and the fcc-based clusters of like sizes, 309 and 314, compared to the free-beam measurement by H\"ovel \textit{et al.}~\cite{hovel-93}, which shows the good agreement for the fcc cluster, whereas the icosahedral clusters have a higher plasmon energy.
 \label{fig_ico_fcc}
}
\end{figure*}
%%%%%%%%%%%%%%%%%%%%%%%%%%%%%%%%%%%%%%%%%%%%%%%%%%%%%%%

%%%%%%%%%%%%%%%%%%%%%%%%%%%%%%%%%%%%%%%%%%%%%%%%%%%%%%%
\begin{figure*}
\centering
\includegraphics[width=0.85\textwidth]{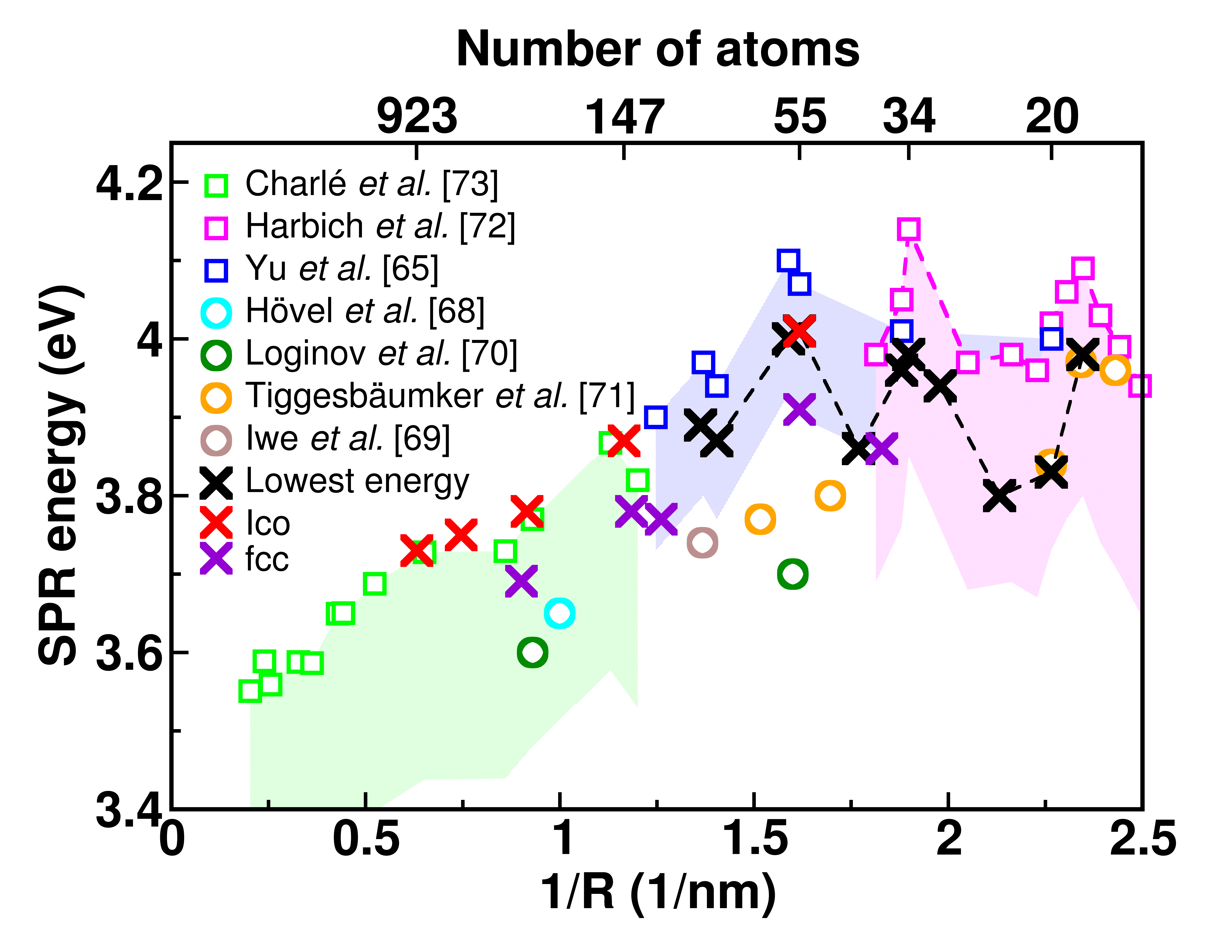}
\caption{Size-dependence of the LSPR in Ag clusters. Our calculated TDDFT+U results are represented with three differently colored crosses: black for the theoretically found lowest-energy clusters (shown in Figure \ref{fig_spectra}), red for the icosahedral clusters (shown in Figure \ref{fig_ico_fcc}(a)), and violet for the fcc-based clusters (shown in Figure \ref{fig_ico_fcc}(b)). Open circles are the experimental values measured in the free-beam experiments~\cite{iwe2023metal,loginov2011photoabsorption,hovel-93,tiggesbaumker1993blue} whereas open squares are the experimental points measured on clusters embedded in neon or argon matrices ~\cite{yu2018optical,harbich1993optical,charle1984optical}. To compensate for the matrix effect of solid neon and argon, the corresponding points were blue shifted by 0.17\,eV and 0.29\,eV, respectively. The pink, blue, and green shaded regions indicate the range between the unshifted and the shifted results. For a detailed discussion of these shifts, please refer to section 1 of the SI. Consistent with the curves shown in Fig.~\ref{fig_ico_fcc}, the fcc clusters have lower plasmon energies than the icosahedral clusters, lying close to or slightly above the range defined by the free-beam experiments. The small and intermediate clusters from Fig.~\ref{fig_spectra} show clear oscillation-like variations due to electronic shell-closing effects, as highlighted by the dotted black lines.
  \label{fig_energies}
  }
\end{figure*}
%%%%%%%%%%%%%%%%%%%%%%%%%%%%%%%%%%%%%%%%%%%%%%%%%%%%%%%

We consider a range of clusters and nanoparticles representative of the clusters studied in the available literature. It comprises (i) small to medium-sized clusters, the optimized structures of which correspond to the geometries used in Refs.~\cite{anak2014time} and ~\cite{schira2019localized} in order to ensure meaningful comparison with the previous calculations. These structures correspond mostly to those found by Chen \textit{et al.}~\cite{chen-13} and are, presumably, the lowest-energy structures of their respective sizes. Furthermore, we consider icosahedral structures as found in many experiments. In particular, the icosahedral structure has been found to be the most stable geometry for the Ag$_{55}$ cluster~\cite{schooss2005structures}. These icosahedral clusters are nearly spherical and particularly popular in theoretical studies because they allow for a series of sizes (13, 55, 147, 309, 561, 923... atoms) without any change of symmetry or morphology, and there are no questions as to how to cut the clusters out of the bulk fcc lattice (facets...) Nonetheless, while small icosahedra are both expected and also found in experiments~\cite{schooss2005structures}, the five-fold symmetries cannot exist in the bulk material and the icosahedra become more and more strained with size. This induces a crossover \cite{baletto-02}, where larger clusters exhibit a preference for fcc-based structures. Consequently, in addition to the icosahedra, we consider a number of nearly spherical fcc-based clusters, i.e., clusters cut out of the fcc bulk lattice.
\\

The calculated results are shown in Figs.~\ref{fig_spectra},\ref{fig_ico_fcc}, and \ref{fig_energies} along with available experimental results. The experimental spectra of the smaller, mostly non-spherical clusters shown in Fig.~\ref{fig_spectra} have been obtained in different experiments on size-selected clusters in rare-gas matrices at low temperatures. They have been taken directly from Refs.~\cite{yu2018optical} and \cite{lecoultre2011ultraviolet}, and they contain an experimental intricacy: the respective rare-gas matrices are generally assumed to induce a dielectric red-shift of the spectra, which means in turn that in order to compare between experiments in different rare gases or with clusters in vacuum, these shifts must be ``corrected''. The details of these shifts have been diversely discussed in the literature in the past~\cite{fedrigo-92,harbich1993optical,lecoultre2011ultraviolet,gervais}. For instance, while for the intermediate sizes between 20 and 100 atoms, shifts of around 0.17\,eV for Ne and 0.29\,eV for Ar have been suggested and applied by different authors, they do not seem to apply for the smallest clusters~\cite{gervais,yu2018optical,charle1984optical}. Our work does not intend to resolve this issue, but we provide in the supplementary material section 1 a more extensive discussion of the subject and on the question as to how it impacts the comparison with our calculated data. We show, in particular, that on the energy scale of Fig.~\ref{fig_spectra}, the good agreement does not depend on the details of these shifts.
  \\

  The spectra of the smallest clusters in Fig.~\ref{fig_spectra} contain multiple peaks mostly reflecting electron-hole-type excitations. The agreement of the calculations with these spectra is very good, the multiple peaks (at least below 5\,eV) are well reproduced. Starting from about 20 atoms all the way upto 92 atoms, a clear broad LSPR band arises in the absorption spectra, with peak energies lying within the range of 3.80 and 4.01 eV. However, we have not yet entered the scalable size range, as the size-dependence is not yet smooth and monotonous. Deviations from a smooth size-dependence originate from electronic or atomic shell-closings as discussed below, as well as from deviations of the clusters from spherical symmetry. Such deviations give rise to both energy shifts and the occurrence of additional peaks and shoulders in the spectra. This is exemplified for the case of Ag$_{58}$ in the supplementary material, Fig.~S7. The geometry of the cluster corresponds to the icosahedral Ag$_{55}$ structure, with three more atoms bonded to the same facet~\cite{chen2013prediction}. While the Ag$_{55}$ structure is very close to spherical, the Ag$_{58}$ presents an elongated cluster. Such shapes have different resonance energies along the three different axes~\cite{baishya2008optical,grigoryan2013optical} which leads to additional peaks or shoulders in the spectra as seen very clearly in both the calculated and the measured spectrum of the Ag$_{58}$. The overall effect of the elongation is a red-shift of the spectrum~\cite{baishya2008optical}. This effect is similar to the purely classical aspect-ratio dependence in elongated nanorods~\cite{lopez2014aspect,sinha2017classical}. Broadening effects can then determine how clearly the different peaks are visible in the overall spectrum. 
For the spectra of these small and medium-sized clusters, the agreement between experiment and calculation is good, the remaining differences are of the order of 0.1\,eV (cf.\ the discussion in the SI). The comparison with previous calculations is shown in the supplementary Figs. S4 and S5.
  \\

  The spectra of the icosahedra and of the fcc-based larger clusters are shown in Fig.~\ref{fig_ico_fcc}. These larger clusters all show the expected broad, smooth LSPR band. In the case of the icosahedral clusters, the size-dependence of the LSPR energy is clearly monotonous, which is not the case for the fcc-based structures. We note that for the size of around 300 atoms, the spectrum of the 314-atom fcc cluster is in very close agreement with the measured gas-phase spectrum of Hövel \textit{et al.}~\cite{hovel-93} (see inset of Fig~\ref{fig_ico_fcc}), whereas the icosahedral cluster of essentially the same size, 309 atoms, has a higher energy.

  In general, all the fcc-based clusters that we considered were found to have LSPR energies lower than the icosahedral ones. A direct comparison of the calculated spectra between icosahedral and fcc-based clusters of similar sizes is provided in the supplementary information (see supplementary Fig. S9). In this comparison, the fcc-based structures are approximately 0.1 eV lower in energy. For the size of about 150 atoms, this difference had already been pointed out previously but not further discussed~\cite{weissker-metal-clusters}, and it is consistent with the classical calculations of Ref.~\cite{noguez2007surface}. 
  \\

  In Fig.~\ref{fig_energies}, we show the LSPR energies as a function of inverse radius for the clusters which have an identifiable LSPR. This permits an overview of the different results and clearly brings out the size-dependence of the plasmons. The experimental results are measured in two different types of experiments: in gas-phase (free-beam) experiments or on clusters embedded in rare-gas matrices as described above. The two types of experiments are consistent among themselves, but differ from each other: the free-beam energies lie consistently below the shifted measurements on the rare-gas-embedded clusters, which renews the questions about the matrix-induced shifts. In particular, it demonstrates immediately the problem in the comparison with the experimental points that have been obtained using the two types of measurements. For that reason, we present the results of the rare-gas-embedded clusters in Fig.~\ref{fig_energies} by the shaded bands ranging from the shifted results to the unshifted (as measured) results. 

The measurements of clusters in free-beam experiments do not contain the type of matrix-induced shifts as mentioned above for the clusters in rare-gas matrices. However, a number of effects are expected to lead to small red-shifts compared to the situation described in the calculations. We discuss these effects in the supplementary material section 2.
\\

The calculated results of the fcc clusters and the small clusters lie close to or slightly above (up to about 0.2\,eV) the band defined by the different free-beam results. If the TDDFT+U results are correct, this is precisely what one would expect because a number of effects lead to small red-shifts in the free-beam experiment compared to the ideal situation of the calculations. These effects include size-distributions, temperature, negative charges, and the presence of a surrounding helium droplet in some cases~\cite{loginov2011photoabsorption}. They are discussed in detail in supplementary material section 2. By contrast, the series of icosahedral clusters produces higher energies.
The difference with respect to the fcc clusters is interesting, in particular because the icosahedra have then been used in a large variety of theoretical studies by many groups, see for instance Refs.\ \cite{weissker-metal-clusters,lopez-bimetal-13,kuisma2015localized,iida-14,rossi-17-KSdecomp,koval-16,baseggio-16}.

Below about 100 atoms, we enter the non-scalable size regime where ``each atom counts'' and shell closings influence the size dependence strongly, as clearly seen in the measurements of Harbich \textit{et al.}~\cite{harbich1993optical,yu2018optical}. This effect is apparent in our calculations involving the lowest-energy structures, shown by the black crosses connected with black dashed lines. In addition to the points extracted from the spectra shown in Figure~\ref{fig_spectra}, further calculations were performed, their spectra are provided in the SI (see supplementary figure S2). Clearly, the three structures Ag$_{18}$, Ag$_{34}$ and Ag$_{92}$ where closed-shell configurations are expected show maximum values of the plasmon energy vs.\ inverse size. Somewhat exceptionally, the plasmon of Ag$_{58}$, which also has a closed shell electronic configuration (cf.\ supplementary figure S8), was found to be slightly lower in energy than that of the highly symmetric Ag$_{55}$ cluster, which can be attributed at least in part to its elongated shape (see supplementary figure S5). In addition, this is a case where the electronic shell closing and the structural shell closing (at 55 atoms) interfere with each other. This effect has already been mentioned in Ref.~\cite{yu2018optical}.

We note that the absolute energies do not conincide particularly well with the shifted published results of the rare-gas-embedded clusters --- our calculated energies lie within the band defined by the shifted and the unshifted energies. However, the \textit{variations}, with their maxima determined by the shell-closings, are well reproduced.
In other words, while the question of the absolute values of the plasmon energies and the treatment of the matrix-related shifts remains, our calculations reproduce the effect of the shell-closings on the plasmon energies well.
\\

\textbf{Hypothetical size dependence of the effective $U$ value.}
We have, until now, used the constant value of 4\,eV for the effective U, thus assuming it could be used for all sizes. This has been, however, a hypothesis. For instance, absent precise knowledge of the measured structures, one could conjecture that the difference in the LSPR energies between the icosahedral clusters and the free-beam measurements could also be corrected by using a different effective $U$ value. To study the validity of this hypothesis, we re-calculated the spectra with different effective $U$ values for the two smallest icosahedral clusters considered in this work (see supplementary figure S10 and S11). We found that to match the range defined by the free-beam experiments, an effective $U$ value of $\sim$1\,eV and $\sim$\,2.5 would be needed for the Ag$_{55}$ and Ag$_{147}$ clusters, respectively. While this could hypothetically be consistent with a behavior that approaches the bulk limit of 4\,eV rather fast, it is not consistent with our results for the smallest clusters shown in supplementary figure S12. Here, the agreement with experiment degrades clearly when the $U$ value is reduced. In particular, the energies decrease and lie increasingly below even the unshifted rare-gas embedded experimental results. Hence, the fixed value of $U$ = 4\,eV appears to be optimal, and any size-dependence of $U$ is expected to be very weak. This, in turn, is an important result in its own right because it demonstrates a surprising transferability of the U: the value determined for bulk silver can be used for all cluster sizes. A self-consistent determination of the $U$ parameters adapted finely to different situations and also to the different atoms in a given system would be desirable. We undertook such calculations using the ACBN0 functional with the octopus code, but the calculations were unsuccessful and returned unphysical results for unknown reasons
\\

%%%%%%%%%%%%%%%%%%%%%%%%%%%%%%%%%%%%%%%%%%%%%%%%%%%%%%%
%%%%%%%%%%%%%%%%%%%%%%%%%%%%%%%%%%%%%%%%%%%%%%%%%%%%%%%

\textbf{Comparison with previous calculations.} For larger clusters, as mentioned above, local and semilocal functionals like the simple local-density approximation (LDA), cf.\ Ref.~\cite{iida-14}, and different generalized-gradient approximations (GGA) like the PBE functional~\cite{perdew1996generalized}, the Wu-Cohen~\cite{WC-functional} functional, and the asymptotically corrected LB94 functional~\cite{LB94} have been the most widely used approximations until now~\cite{malola-13,sinha-roy-23,koval-16,aikens-08,burgess-14}. The meta-GGA functional VS98 has been used for larger gold clusters~\cite{sakthivel-20-VS98}. 
The use of the local and semilocal functionals results, in general, in the filled d shells being positioned too close to the Fermi energy and, due to the resulting overestimation of d screening of the LSPR, underestimates LSPR energies (see supplementary figure S3)~\cite{weissker2011optical,weissker2014optical,sinha-roy-23}. 
For larger clusters, Kuisma \textit{et al.}\ have employed the Gritsenko-van Leeuwen-van Lenthe–Baerends solid-correlation potential (GLLB-SC)~\cite{kuisma-10,GLLB-functional} (see supplementary figure S5). Their results~\cite{kuisma2015localized} are only slightly higher than ours for the larger clusters, but the quality of these results seems to degrade strongly with decreasing size: there is a clear overestimate of the LSPR in Ag$_{55}$ with respect to all available experimental results shown in Fig.~\ref{fig_energies}.

%%%%%%%%%%%%%%%%%%%%%%%%%%%%%%%%%%%%%%%%%%%%%%%%%%%%%%%

For small clusters, excellent results have been obtained using the numerically costly range-separated hybrid functionals~\cite{schira2019localized,rabilloud2013assessment} (see figure S4). In general, the quality of our spectra is comparable with those results. However, interestingly, the largest cluster that seems to have been published using this method, Ag$_{147}$, has an LSPR energy that is distinctly higher than that obtained with TDDFT+U, and in view of Fig.~\ref{fig_energies} it seems decidedly too high compared to all available experiments. In addition, the calculations using the range-separated hybrid functionals would not be feasible with today's numerical means for the larger clusters, beyond maybe 200 silver atoms.

In comparison with these calculations, only the TDDFT+U method yields reliable spectra over the full treatable size range, from the smallest clusters to the 3\,nm nanoparticle of 923-atoms.
\\
\\
\textbf{Effect of $U$ correction on localization and density of states.} While the spectra shown and discussed above are the central result of our present work, we need to analyze the effect of the $U$ correction on the other observables as well in order to obtain a full understanding of our results.

\begin{figure*}[t]
  \centering
  \includegraphics[width=0.9\textwidth]{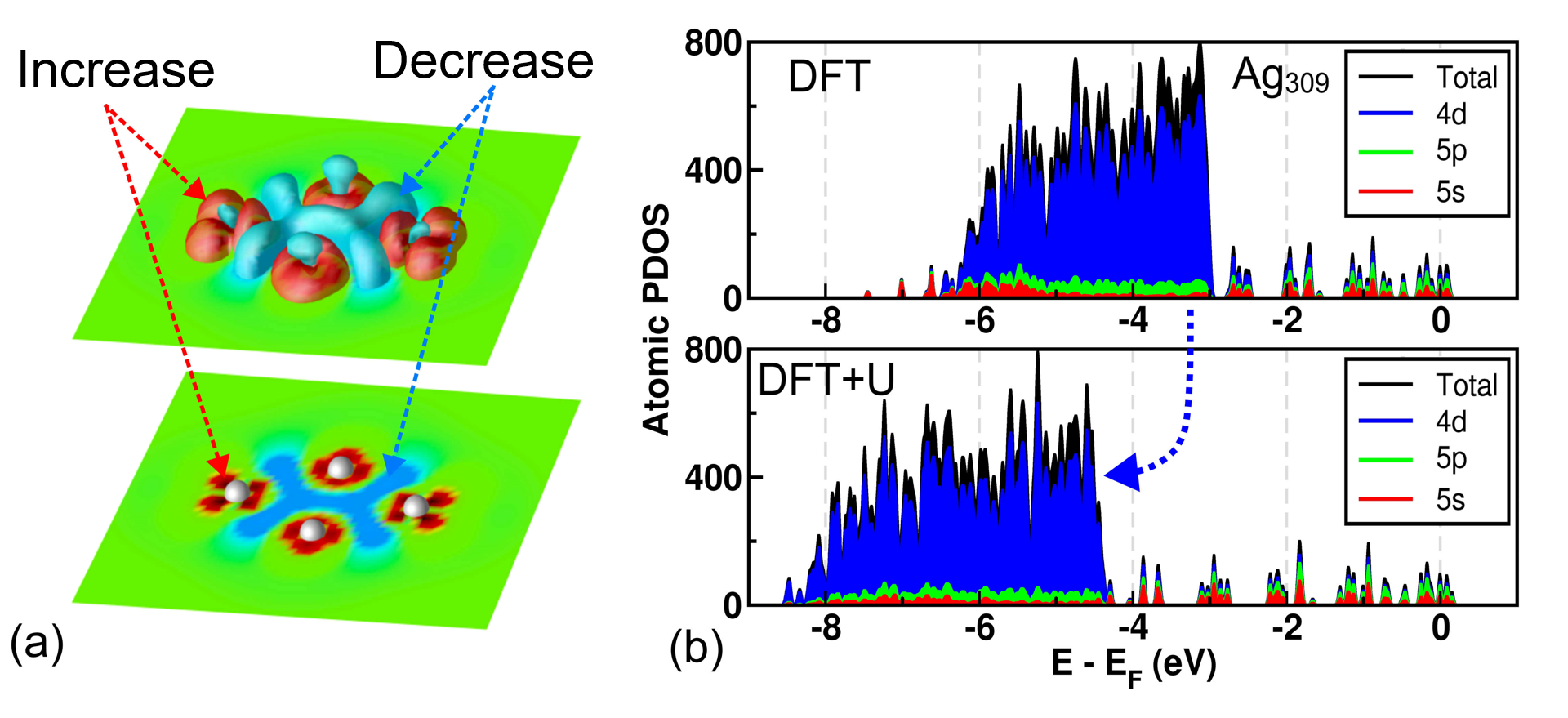}
  \caption{(a) The color-coded slab represents the difference in total electronic charge density obtained using DFT+U and DFT for the Ag$_4$ cluster. In this representation, red and blue colors indicate regions of electron density increase and decrease, respectively, thus showing the increased localization. (b) Projected density of states (PDOS) of Ag$_{309}$, calculated using DFT (top panel) and DFT+U (bottom panel) method. The shift of the d band to lower energies is clearly seen. A comparison with experiment is shown in Supplementary Figure S6.}
%  \label{fig_3}
  \label{fig_density_and_dos}
\end{figure*}

In particular, the redistribution of the total electronic density is shown in Figure~\ref{fig_density_and_dos}, presented as the difference between the total charge density of the Ag$_4$ cluster calculated using GGA (PBE) with inclusion of the Hubbard correction of 4\,eV and GGA (PBE) without the correction. This difference is shown on a color-coded slab, showing regions of accumulation and depletion of electrons in red and blue, respectively. The region of accumulation exhibits the characteristic shape of the d-orbitals, visually confirming the enhanced localization of the 4d electrons upon applying the $U$ correction term. 
%%%%%%%%%%%%%%%%%%%%%%%%%%%%%%%%%%%%%%%%%%%%%%%%%%%%%%%
%%%%%%%%%%%%%%%%%%%%%%%%%%%%%%%%%%%%%%%%%%%%%%%%%%%%%%%

%%%%%%%%%%%%%%%%%%%%%%%%%%%%%%%%%%%%%%%%%%%%%%%%%%%%%%%
\begin{figure*}[t]
  \centering
  \includegraphics[width=0.9\textwidth]{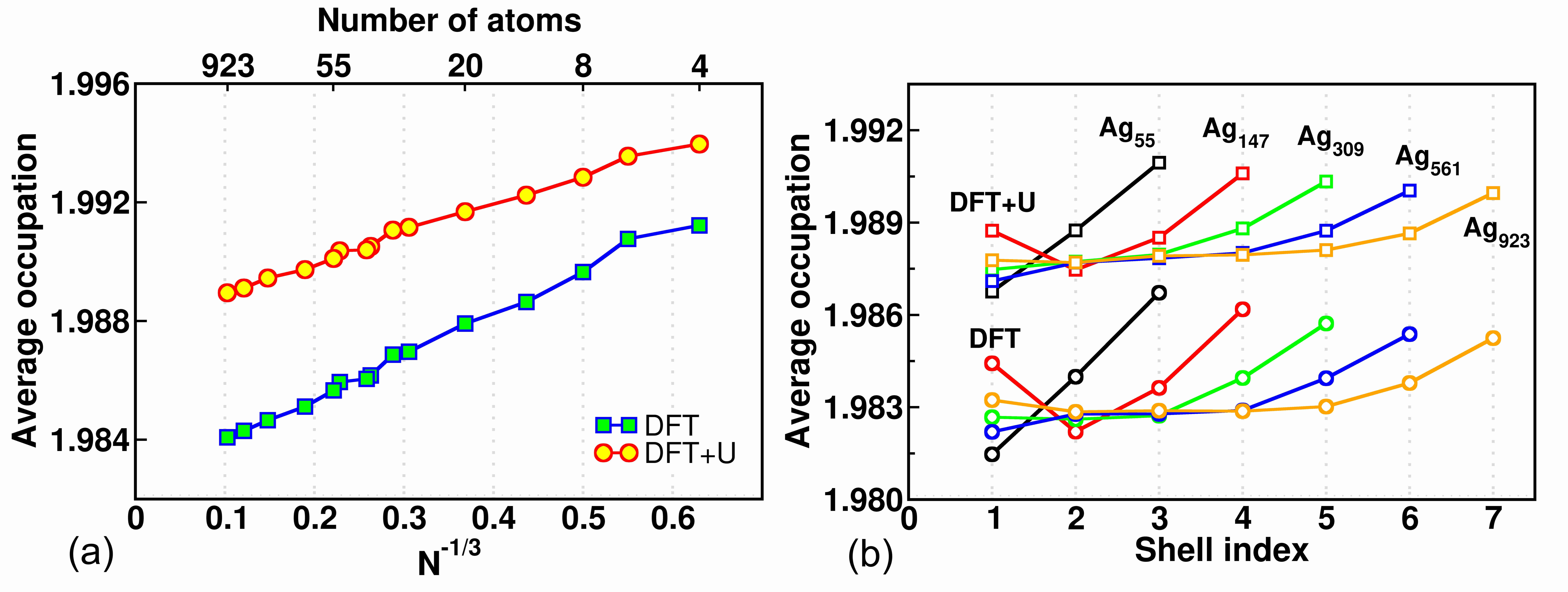}
  \caption{(a) Average occupation of 4d-orbitals plotted against the inverse cube root of the number of particles in a cluster, illustrating two effects: higher average localization of d electrons when the $U$ correction is introduced, and enhanced average localization in smaller clusters compared to larger ones. (b) Shell-wise average occupation of the d orbitals for five icosahedral structures, showing increased localization of the d electrons in the outer shells, indicating weaker Coulombic screening on the surface.}
 \label{fig:localization}
\end{figure*}
%%%%%%%%%%%%%%%%%%%%%%%%%%%%%%%%%%%%%%%%%%%%%%%%%%%%%%%

Additionally, the Hubbard $U$ correction shifts the 4d states to lower energies with respect to the Fermi energy, as it is evident in the projected density of states (PDOS) of the Ag$_{309}$ cluster, shown in Figure \ref{fig_density_and_dos}(b). The correction in the energetic position of the 4d states ensures that the threshold of the interband transitions from the occupied 4d states into higher unoccupied states beyond the Fermi energy appears at the correct energy.

A comparison with experimental photoemission spectra is shown in Supplementary Fig. S4 shows that the d-band edge is well corrected, even though the width of the d band is somewhat overestimated. The latter point does not, however, impact the calculation of the spectra and plasmon energies strongly, as is evidenced by the results shown above.
In addition, our calculation captures all the important features of the occupied states above the d band present in the UV photoemission spectra of Ag$_{55}$. This shows that, as expected, the main effect of the inclusion of the $U$ correction is to correct the principal shortcoming of the simple functionals --- namely, the incorrect description of the localized d states, with the d band lying too close to the Fermi energy and the interband transitions appearing too low in the spectra, thereby interfering unphysically with the LSPR and overestimating the screening of it.~\cite{cottancin2006optical,PRL-serra-rubio-97,sinha-roy-23}. Thus the correction in turn significantly improves the agreement with experimental optical spectra as shown above.
\\

To quantify the localization effect of the d-orbitals, we calculated the average occupation of the d-orbitals for clusters of all sizes with and without the Hubbard $U$ correction. Our analysis found that the average occupation of the d orbitals in DFT+U calculations was higher in comparison to the calculations without $U$ correction, implying enhanced d electron localization in agreement with the results shown in Fig.~\ref{fig_density_and_dos}.  Additionally, we observe an increase in the \textit{average} occupation of the d orbitals with decreasing cluster size. 

To gain a deeper understanding of how the different atomic sites contribute towards the average localization of the 4d electrons, we analyzed the average occupation of the 4d orbitals shell-wise in the five icosahedral structures, which is shown in figure \ref{fig:localization}(b). It is evident from the figure that the d electrons are more localized on the two outermost shells than inside in all cases. Below the two outermost layers, the localization is roughly constant. This means that the increase of the average localization in Fig.~\ref{fig:localization}(a) is due to the increasing surface-to-volume ratio for decreasing size.

%%%%%%%%%%%%%%%%%%%%%%%%%%%%%%%%%%%%%%%%%%%%%%%%%
All these results are coherent with our general understanding of the d screening of the LSPR. An increase of the localization implies a reduction of the polarizability of the d electrons, thereby reducing their screening effect as discussed above. The increased average localization in DFT+U thus provides a contribution to the blue shift of the plasmon compared to GGA (PBE). The increase of average localization with decreasing cluster size and the resulting decrease of the average d screening are consistent with the observed blue shift in the LSPR.

It is interesting to note that the increased localization of the outermost atoms' d states, which implies their decreased polarizability and, hence, a decreased contribution to the overall screening at the surface, points in the same direction as the above-mentioned effect of the surface layer of reduced d screening due to the spatial localization of the d electrons around the atom cores. This is even more relevant as it is known that the outermost atomic shell plays an important role in the determination of the optical properties~\cite{lopez-bimetal-13}.

The model of a reduced screening layer generally applied in many jellium calculations~\cite{liebsch-93,tiggesbaumker1993blue,campos2019plasmonic} is, therefore, an effective model that is able to represent both the localization of the d electrons at a short distance from the surface \textit{and} the increased localization around each atom in the surface layer.

%%%%%%%%%%%%%%%%%%%%%%%%%%%%%%%%%%%%%%%%%%%%%%%%%%%%%%%
\section{Conclusion}\label{sec3}
%%%%%%%%%%%%%%%%%%%%%%%%%%%%%%%%%%%%%%%%%%%%%%%%%%%%%%%

In conclusion, the TDDFT+U method provides optical spectra over the full size range from few-atom silver clusters to nanoparticles of about 3\,nm in diameter, corresponding to about one thousand atoms and containing $\approx$10,000 active electrons. In addition to the electron-hole-type transitions of small clusters, our calculations obtain the broad plasmon resonances of the large spherical clusters and their size-dependence. The numerical effort is only slightly larger than that of pure TDDFT calculations.

Precise comparison with experiment is complicated by inconsistencies in the experimental literature and uncertainties about the cluster structures. Our results for small clusters up to about 100 atoms reproduce the spectra measured on rare-gas-embedded clusters very well, including in particular the spectra with multiple peaks. Their calculated plasmon energies reproduce in particular the oscillation-like behavior due to electronic shell-closing effects, whereas comparison of the absolute energies is only possible up to remaining differences of up to about 0.2\,eV due to uncertainties in the treatment of the  matrix shifts needed in the comparison. By contrast, the calculated plasmon energies of fcc-based nearly spherical clusters are in good general agreement with, i.e., close to or slightly above, the range defined by available free-beam experiments. The series of icosahedral clusters has generally higher plasmon energies.

The TDDFT+U approach is the only method that, at this time, achieves this degree of agreement with experiment over the full size range because the costly range-separated hybrid functionals cannot realize calculations of clusters beyond $\approx$ 200 atoms, and the solid-state-derived meta-GGA calculations that fare well for large clusters seem to fare poorly below about 100 atoms. 

The value of the effective $U$ turns out to be surprisingly transferable for the silver clusters. The same value of 4\,eV that had produced spectra in good agreement with experiment for bulk silver was used without any adaptation for all the clusters in the present work, including the smallest one, Ag$_4$. Tests with different $U$ values indicate that any size-dependent variation of $U$ will be weak.

To use the full power of the TDDFT+U  method, a more precise comparison with experiment would be desirable. We hope that our work can motivate further experimental investigations that will consolidate the available results and, in particular, help settle the open questions about cluster-matrix interactions and the resulting shifts.

Clearly, the TDDFT+U method can likewise be used to calculate systems where the clusters are coupled to each other or to bio-molecules, DNA strands, etc. Our results open the pathway to direct TDDFT+U calculations of many practically relevant systems and processes, including, for instance, medical imaging applications, biomolecule labeling, sensing, and many others.

%%%%%%%%%%%%%%%%%%%%%%%%%%%%%%%%%%%%%%%%%%%%%%%%%%%%%%%
\section*{Methods}
%%%%%%%%%%%%%%%%%%%%%%%%%%%%%%%%%%%%%%%%%%%%%%%%%%%%%%%

\textbf{Experimental spectra and LSPR energies:} Experimental absorption spectra of the clusters with 20 to 92 Ag atoms were scanned from Ref.~\cite{schira2019localized} where they include already a blue shift of 0.17\,eV to compensate for the dielectric effect of the neon matrix present in the original experiment~\cite{yu2018optical}. The absorption spectra of the smallest clusters embedded in rare-gas neon matrix were taken directly from Ref.~\cite{lecoultre2011ultraviolet} without any matrix correction, as it was done in previous publications~\cite{rabilloud2013assessment,anak2014time,lecoultre2011ultraviolet}.

For the plot of the LSPR energy vs.\ inverse radius, Fig.~\ref{fig_energies}, the radius of the particle was approximated using $R=r_s*N^{1/3}$\r{A}, with $r_s=1.626$ being the electronic density parameter of bulk silver and N being the number of Ag atoms. The values of Charlé \textit{et al.} and Harbich \textit{et al.}, were scanned from Refs.~\cite{charle1984optical,haberland-13} and consequently contain a blue-shift of 0.29\,eV to compensate the dielectric shift of the Ar matrix discussed in that reference~\cite{haberland-13}.  The data points from Yu \textit{et al.}~\cite{yu2018optical} were taken from the table in that reference and were blue shifted by 0.17\,eV by us to compensate for the dielectric shift of the Ne matrix.
The values of Refs.~\cite{loginov2011photoabsorption}, \cite{hovel-93}, and \cite{tiggesbaumker1993blue} were taken directly from the references, naturally with no shifts. For a detailed discussion of the shifts, please refer to the SI, section 1.
\\

\noindent
\textbf{Geometries:} The structures used for the calculations shown in Figure~\ref{fig_spectra} were taken from the works of Schira \textit{et al.}~\cite{schira2019localized} and Anak \textit{et al.}~\cite{anak2014time}. Additional structures, which were primarily used to calculate more points for the $E_{\rm LSPR}$-vs.-$1/R$ curve, were taken from the work of Chen \textit{et al.}~\cite{chen2013prediction}. The Ag$_{34}$ structure was constructed by selectively removing one atom from the Ag$_{35}$ cluster, ensuring the overall shape is roughly spherical. This newly constructed structure when fully relaxed with the VASP code was found to be only 1.3\,meV/atom higher in energy than the structure for the same cluster provided by Chen \textit{et al.} The icosahedral and fcc-based geometries were constructed by us. For consistency, all the structures were again relaxed with the VASP code~\cite{kresse1996efficiency,kresse1999ultrasoft}, using the GGA functional as parameterized by Perdew and Wang for the exchange and correlation~\cite{PW91}.
\\	

\noindent
\textbf{Calculations:} Both the ground state and the optical absorption spectra of all structures were calculated using the DFT+U method as implemented~\cite{tancogne2017self} in the \texttt{octopus} code~\cite{codetancogne2020octopus}. Our investigation uses the rotationally invariant formulation\cite{dudarev1998electron} of the DFT+U method where $U_{\rm eff}=U-J$ is used but generally referred to as just U. It acts locally on the 4d orbitals at all Ag atomic sites. Absorption spectra were calculated using real-time TDDFT+U. We use the $U$ value of 4.0\,eV for all calculations. This choice was motivated by the finding that this value provides good dielectric functions for bulk silver~\cite{avakyan2020theoretical}. Apart from the $U$ correction, we used PBE to approximate the exchange-correlation functional~\cite{perdew1996generalized}. The interactions between the electrons and the ions were described using norm-conserving Troullier–Martins pseudopotentials~\cite{troullier1991efficient}, treating 11 valence electrons corresponding to the 10 4d electrons and one 5s electron per Ag atom explicitly. For all the spectra shown in Figure~\ref{fig_spectra}, the grid spacing and the radius of simulation was set to 0.18~\r{A} and 7.5~\r{A}, respectively. This is the so-called ``minimum" radius to define the simulation box in octopus. It indicates the radius of spheres around each atom, the superposition of which makes up the domain used for the calculation. For Ag$_{147}$ to Ag$_{561}$ icosahedral clusters, a grid spacing of 0.20~\r{A} and  a simulation radius of 7.5~\r{A} were used, while for Ag$_{923}$, a smaller radius of 5.0~\r{A} was used. This is justified because the differences in the LSPR energetic position due to the use of smaller radius (5.0~\r{A}) decrease strongly with increasing cluster size in our tests, decreasing from 0.05\,eV for Ag$_{55}$ to 0.01\,eV for Ag$_{561}$, and thus being negligible for Ag$_{923}$. For the fcc-based clusters the spacing and the radius were set to 0.20 \r{A} and 7.5 \r{A}. To calculate the absorption spectrum, we have used the Yabana-Bertsch time-evolution formalism~\cite{yabana1996time}, which involves real time propagation of the wavefunctions after a delta kick at $t=0$. As a propagator, we have used Approximated Enforced Time-Reversal Symmetry (aetrs), with a time step of $\approx$ 0.0016 fs (0.0024 $\hbar/eV$) and a total propagation time of $\approx$ 26 fs (40 $\hbar/eV$). For the icosahedral and fcc-clusters, a shorter propagation time of $\approx$ 13 fs (20 $\hbar/eV$) was used.

The LSPR energies for all the structures were identified by limiting the evolution time to $\approx$ 13 fs (20 $\hbar/eV$), which is equivalent to applying a larger broadening to the spectra. The plasmon energy was then identified as the maximum of the peak.
\\

For calculating the average localization of the d-orbitals, we used the following expression 
\begin{equation}
  \langle n_{4d}\rangle=\frac{1}{I}\sum_{j}^{I}\frac{{\rm trace}(n_{m,m'}^j)}{5} \label{avg_occupation}
\end{equation}
where $I$ is the total number of atomic sites,  $n_{m,m'}^j$ are the occupation matrices defined as the density matrix of a localized orbital atomic basis set $\{\phi_{j,m}\}$ obtained from the pseudopotential, attached to the $j_{th}$ atomic site~\cite{tancogne2017self}. From the Kohn-Sham wavefunctions, occupation matrix for the $j_{th}$ atom is given by
\begin{equation}
  n_{m,m'}^j=\sum_{n}f_n <\psi_n|\phi_{j,m}><\phi_{j,m'}|\psi_n> \label{occupation_matrix}
\end{equation}
where $f$ is the occupation of the $n_{th}$ KS state and $m$ or $m'$ are the angular quantum numbers of the localized atomic basis set, which in our case is restricted to the 4d orbitals. For calculating the shell-wise average occupation of the d orbitals for the icosahedral structures, we used the same eq (\ref{avg_occupation}), taking averages over the atomic sites of one shell at a time. The first shell refers to the central atom alone.

\backmatter

\bmhead{Acknowledgments}

The authors thank Franck Rabillould for providing the geometries of the smaller clusters, which enabled the detailed comparisons in the present work. Furthermore, the authors thank Jean Lermé, Matthias Hillenkamp, Gérard, Franck Rabilloud, and Lucia Reining for enlightening discussions. Moreover, the authors acknowledge the contribution of the International Research Network IRN Nanoalloys (CNRS). Mohit C.\ thanks ED352 of Aix-Marseille University for the PhD scholarship.

\section*{Additional information}

\textbf{Supplementary Information} accompanies this paper. It contains a detailed discussion of the matrix-related shifts of the plasmon energies as well as of the influences that may modify plasmon energies in the free-beam experiments. It contains further additional spectra, a comparison with previous GGA results, as well as with results of range-separated hybrid functionals and the GLLB-SC functional. We show a rough comparison with experimental photo-emission spectra, demonstrate the effect of deviations from spherical clusters, and highlight the differences between the spectra of icosahedral and fcc-based clusters. The shell closings of Ag$_{38}$ and Ag$_{38}$ are demonstrated using projected densities of states. Finally, the effect of changing the effective $U$ value is demonstrated for spectra and plasmon energies.
\\

\noindent
\textbf{Competing interests:} The authors declare no competing interests.

%%===================================================%%
%% For presentation purpose, we have included        %%
%% \bigskip command. please ignore this.             %%
%%===================================================%%
\bigskip

%%=============================================%%
%% For submissions to Nature portfolio Journals %%
%% please use the heading ``Extended Data''.   %%
%%=============================================%%

%%=============================================================%%
%% Sample for another appendix section			       %%
%%=============================================================%%

%% \section{Example of another appendix section}\label{secA2}%
%% Appendices may be used for helpful, supporting or essential material that would otherwise 
%% clutter, break up or be distracting to the text. Appendices can consist of sections, figures, 
%% tables and equations etc.

%%===========================================================================================%%
%% If you are submitting to one of the Nature Portfolio journals, using the eJP submission   %%
%% system, please include the references within the manuscript file itself. You may do this  %%
%% by copying the reference list from your .bbl file, paste it into the main manuscript .tex %%
%% file, and delete the associated \verb+\bibliography+ commands.                            %%
%%===========================================================================================%%

%%%\bibliography{}% common bib file
%%% if required, the content of .bbl file can be included here once bbl is generated

%\bibliography{References}

%%\input sn-article.bbl

\end{document}